\documentclass[12pt]{article}
\pdfoutput=1
\usepackage{graphicx}
\usepackage{amssymb}
\usepackage{epsfig}
\usepackage{amsmath}
\usepackage{verbatim}
\usepackage{titlesec}

\csname @addtoreset\endcsname{equation}{section}
\begin{document}
\begin{titlepage}
\title{\bf\Large   Constraints on Natural Supersymmetry from Electroweak Precision Tests  \vspace{18pt}}
\author{\normalsize   Sibo~Zheng$^{1}$ and Yao~Yu$^{2}$ \vspace{15pt}\\
{\it \small $^{1}$Department of Physics, Chongqing University, Chongqing 401331, P.R. China}\\
{\it\small  $^{2}$College of Mathematics and Physics,}\\
{\it\small  Chongqing University of Posts and Telecommunications,
Chongqing 400065, P. R. China}
\\}

\date{}
\maketitle \voffset -.3in \vskip 1.cm \centerline{\bf Abstract}
\vskip .3cm  
The observed Higgs boson mass and the naturalness argument leave us a
narrow window for the soft mass spectrum in natural supersymmetry 
that can be studied through the electroweak precision tests (EWPTs).
We divide the analysis into
the Higgs sector constrained by the charged Higgs mass bound,
the neutralino-chargino sector constrained by the chargino mass bound,
and the third-generation squark sector tightly constrained by the observed Higgs mass.
Total contributions to EWPTs in the MSSM and NMSSM are both presented.
It turns out that natural MSSM is excluded at 68\% CL but consistent at 95\% CL,
whereas natural NMSSM with nearly degenerate conditions is excluded at 68\% CL 
but consistent at 99\% CL for $\lambda\leq 0.6$.

\thispagestyle{empty}

\end{titlepage}
\newpage
\section{Introduction}
With a standard model (SM)-like Higgs with mass $126\pm 1$ GeV discovered \cite{higgsmass}, 
the first run of Large Hadron Collider (LHC) has not observed signals of supersymmetry (SUSY) yet,
but leaves us a few lower bounds on superpartner masses in various simplified models.
These bounds are roughly of order of a few hundred GeVs for sfermions and 1 TeV for gluino.

Despite the absence of SUSY so far, and the fact that the actual Higgs mass requires a large tuning 
in the context of the minimal SUSY standard model (MSSM),
the motivation for SUSY is actually not challenged but strengthened after the discovery of Higgs boson. 
The fine tuning implied by the LHC 2013 data and the fit to Higgs mass can be only relaxed in some subtle SUSY models.

In this paper, we would like to explore natural SUSY through the EWPTs \cite{Peskin}.
The motivation for such study is straightforward given that 
natural SUSY can be efficiently probed thanks to indirect constraints coming from the Z-pole observables, 
top quark and Higgs mass and their total contribution to EWPTs \cite{EWPTlimits}.
The upper bounds on superpartner masses due to the {\it naturalness},
and the lower bounds due to the LHC data give rise to rather narrow window from $\sim$ a few hundred GeVs to $\sim 1$ TeV for natural SUSY soft mass spectrum.
We make use of the EWPTs to explore such a constrained spectrum \cite{Feng, Craig}.
For attempts to address this problem in earlier works, 
see, e.g., \cite{Zheng1, Zheng2, Zheng3, mssmewpt1,mssmewpt2,mssmewpt3,mssmewpt5}.

The paper is organized as follows.
In section 2, we discuss the input parameters and constraints related.
We divide input soft mass parameters into the Higgs sector, 
neutralino-chargino sector and third-generation squark sector.
The input mass parameters of Higgs, neutralino-chargino and third-generation squark sector are mainly constrained 
by the light chargino mass bound \cite{chargedhiggs}, charged Higgs boson bound and Higgs mass fit \cite{higgsmass}, respectively.
We will briefly review the situations in both MSSM and the next-to-minimal supersymmetric model (NMSSM).

In subsection 3.1 to 3.3, we derive the contribution to EWPTs in individual sector numerically,
and discuss the differences between the MSSM and NMSSM.
In subsection 3.4, we combine separate contributions 
to give total estimates for both natural MSSM and NMSSM spectrum.
It is shown that natural MSSM is excluded at 68\% CL but consistent at 95\% CL,
whereas natural NMSSM with nearly degenerate conditions is excluded at 68\% CL 
but consistent at 99\% CL for either $\mu\leq 1$ TeV or $\lambda\leq 0.6$.

Finally, we conclude in section 4.

\section{Input Parameters and Constraints}
In order to show the sensitivity of our analysis to $\tan\beta$,
we will adopt two representative values of $\tan\beta$ as follows,
\begin{eqnarray}{\label{beta}}
{\it \mathbf{MSSM}}&:& \tan\beta =\{10,20\},\nonumber\\
{\it \mathbf{NMSSM}}&: &\tan\beta=\{2,\sqrt{7}\}.
\end{eqnarray}
We consider different values of $\tan\beta$ for the MSSM and the NMSSM 
due to different sensitivity to the Higgs boson mass.
As noticed in the Introduction, we divide the input parameters into those of the scalars of the Higgs sector, 
of the charginos and neutralinos in the so-called neutralino-chargino sector, 
and of stops and sbottoms in the third generation sector.
For other superpartners, including the third generation sleptons and first two-generation sfermions,
their contributions to EWPTs are at least an order of magnitude smaller than what we mentioned above,
and they will be ignored in this study.\\

{\it \underline{Higgs Sector}}\\
The input parameters involved in the Higgs sector of MSSM are,
\begin{eqnarray}{\label{higgs}}
\{{\it m^{2}_{H_{u}},~\mu}\},
\end{eqnarray}
with the soft mass squared $m^{2}_{H_{d}}$ fixed through electroweak symmetry breaking (EWSB) conditions.
For the NMSSM, we choose input parameters as follows,
\begin{eqnarray}{\label{nhiggs}}
\{{\it \mu,~~\lambda,~\kappa,~A_{\lambda}, A_{\kappa}}\},
\end{eqnarray}
with soft mass squared $m^{2}_{H_{u,d}}$ fixed through EWSB conditions. 
Here $\lambda$ denotes singlet-Higgs doublet-Higgs doublet coupling,
$\kappa$ refers to the self coupling of singlet,
and $A_{\lambda}$ and $A_{\kappa}$ are their $A$-terms.
The soft mass parameters are upper bounds from the naturalness 
and lower bounded from direct searches at colliders such as LHC and LEP II.
 
There is a direct constraint on the input parameters of the Higgs sector 
that arises from the lower bound on charged Higgs boson,
$m_{H^{\pm}}\geq 300$ GeV \cite{chargedhiggs} .
We will assume that the Higgs boson is the lightest CP-even state 
and only impose the chargino mass bound when we study this sector individually.
We will impose the Higgs mass when we consider the total contribution in subsection 3.4.
Table 1 summarizes the parameter space of our numerical scan.
Notice that we take a negative up-Higgs mass squared as demanded by EWSB.\\

{\it \underline{Neutralino-Chargino Sector}}\\
The input parameters in the neutralino-chargino sector of MSSM are,
\begin{eqnarray}{\label{neutralino}}
\{{\it M_{1},~ M_{2},~\mu}\}.
\end{eqnarray}
In contrast to the MSSM, additional three parameters $\lambda$, $\kappa$ and the vacuum expectation value of singlet $s$ should be included in the NMSSM. 
The set of input parameters is given by,
\begin{eqnarray}{\label{nneutralino}}
\{{\it M_{1},~ M_{2},~\mu,~\lambda,~\kappa,~s}\}.
\end{eqnarray}

The constraint on $\mu$ in this sector mainly comes from the lower bound on lighter chargino mass, $m_{\chi}>103$ GeV.
Often this mass bound is roughly understood as $\mu >100$ GeV in the literature,
given the explicit dependence of $m_{\chi}$ on $\mu$.
The naturalness upper bounds on wino and bino masses are $\sim 1$ TeV, 
while lower bounds come mainly from direct searches at the LHC 
that constrain the electroweakino masses \cite{electroweakinos}.
For recent review on this topic, see, e.g., \cite{Feng, Craig}.\\

{\it \underline{The third-generation Squark Sector}}\\
The input parameters in the third-generation squark sector of MSSM and NMSSM are the same,
which are given by,
\begin{eqnarray}{\label{squarksector}}
\{{\it m_{\tilde{t}_{L}},~ m_{\tilde{t}_{R}},~m_{\tilde{b}_{L}},~ m_{\tilde{b}_{R}},~A_{t}}\}.
\end{eqnarray}
where $m_{\tilde{t}_{L,R}}$ and  $m_{\tilde{b}_{L,R}}$ refer to squark and sbottom masses, respectively.
$A_{t}$ denotes top-quark $A$-term.

For the MSSM the constraints on these soft mass parameters mainly arise from the fit to Higgs boson mass \cite{higgsmass} reported by the LHC.
In particular, large values for input  mass parameters in Eq.(\ref{squarksector}) are required.
Naturalness upper bounds demand a scale $\sim $ TeV, and for this reason we consider maximal mixing.
In contrast to the MSSM, the fit to Higgs mass mainly depends on $\lambda$ and $\tan\beta$ in the NMSSM,
and the tree-level contribution to Higgs mass can be adjusted large enough.
Conversely, in the NMSSM the dependence on the input parameters of Eq.(\ref{squarksector}) is milder than in the MSSM.

In the light of lower bounds on lighter stop mass \cite{stop1,stop2} and sbottom mass \cite{sbottom1, sbottom2} 
from the LHC experiments,
we choose the lower bounds on soft masses $m_{\tilde{t}_{L,R}}$ and $m_{\tilde{b}_{L,R}}$ as in table 1.

\begin{table}
\begin{center}
\begin{tabular}{|c|}
  \hline
  100~GeV $\leq\mu\leq$ 1000~GeV \\
  -(500 GeV)$^{2}$ $\leq~m^{2}_{H_{u}}\leq$ -(100~GeV)$^{2}$ \\
-100 GeV $\leq A_{\lambda}\leq$ 250 GeV \\
-150 GeV $\leq A_{\kappa}\leq$ 150 GeV \\
0.1 $\leq\lambda\leq$ 0.6 \\
0.05  $\leq\kappa\leq$ 0.6 \\
200 GeV $\leq M_{1}\leq$ 1000 GeV \\
200 GeV  $\leq M_{2}\leq$ 1200 GeV \\
200 GeV  $\leq s \leq$ 2000 GeV \\
550 GeV  $\leq m_{\tilde{t}_{L}},m_{\tilde{b}_{L}}\leq$ 1500 GeV \\
550 GeV  $\leq m_{\tilde{t}_{R}}, m_{\tilde{b}_{R}}\leq$ 1500 GeV \\
100 GeV  $\leq A_{t} \leq$ 3000 GeV \\
\hline
\end{tabular}
\caption{Parameter space for the input parameters.
The upper bounds on soft mass parameters are due to the naturalness argument. 
The lower bound on $\mu$ arises from the lighter chargino mass bound $\geq 103$ GeV.
The upper bound $\lambda\leq 0.7$ follows from the constraint that 
NMSSM stays perturbative up to grand unification (GUT) scale.
Note that this bound can be relaxed in some cases, see, e.g., \cite{Zheng} for recent discussion.
We impose direct constraints on wino and bino masses $M_{1,2}\geq 200$ GeV, 
while we adopt 400 GeV as lower bound on the stop mass parameters.
See the text for explanation on these bounds.}
\end{center}
\end{table}

\section{MSSM vs NMSSM}
\subsection{Higgs Sector}
According to the parameter region chosen as in Table 1,
the contribution to EWPTs in Higgs sector is plotted in Fig.\ref{higgs}.
In this figure, red, blue and yellow region represents deviation to the SM expectation at 68\%, 95\% and 99\% CL, respectively.
Points corresponding to different $\mu$ are shown in different colors.
Fig.\ref{higgs} shows that the maximum values of the $S$ parameter are 0 and 0.017 in the MSSM and NMSSM, respectively. 
The $T$ parameters is always close to zero in both cases.

\begin{figure}
\centering
\begin{minipage}[b]{0.4\textwidth}
\centering
\includegraphics[width=2.5in]{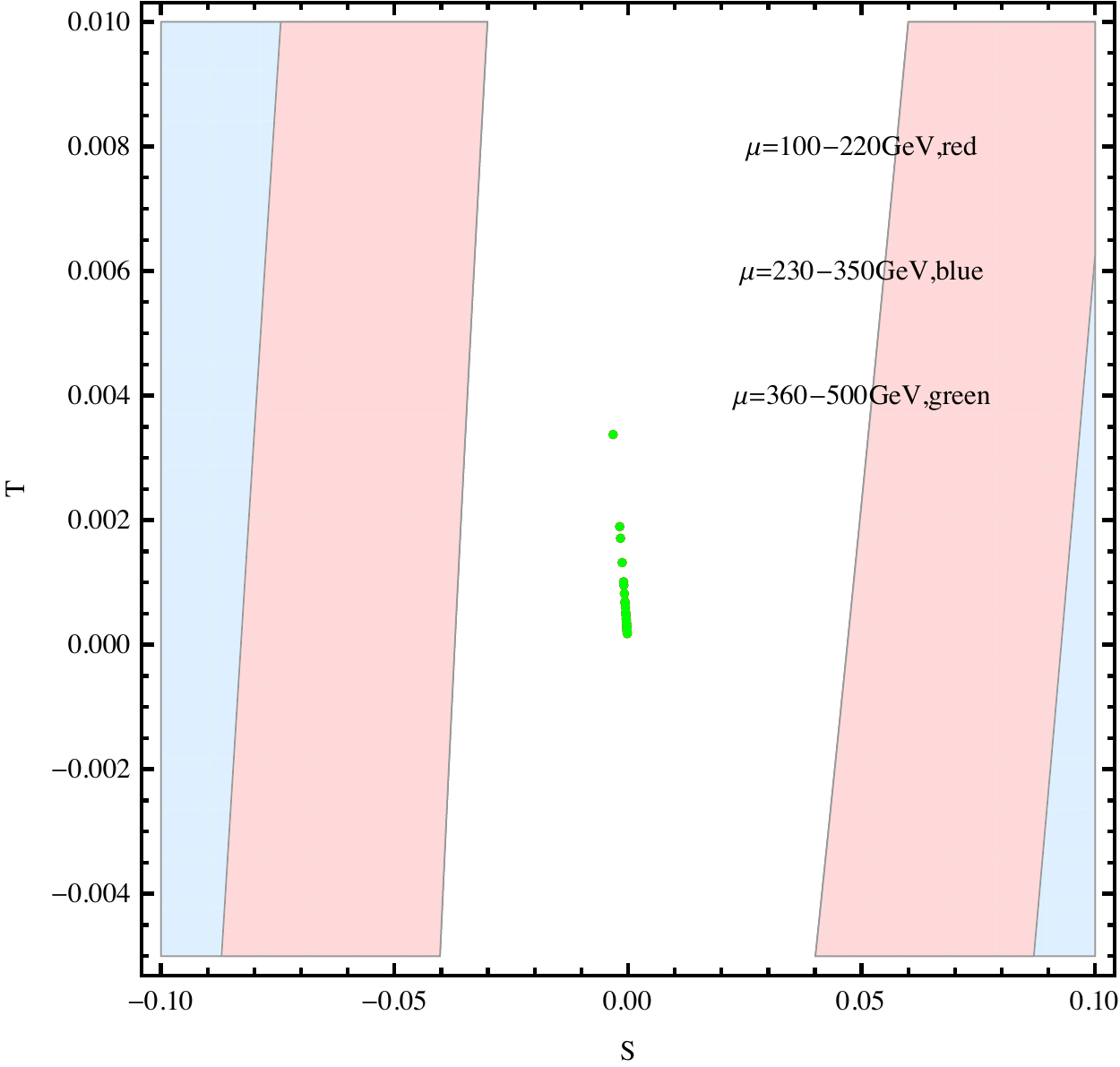}
\end{minipage}%
\centering
\begin{minipage}[b]{0.4\textwidth}
\centering
\includegraphics[width=2.5in]{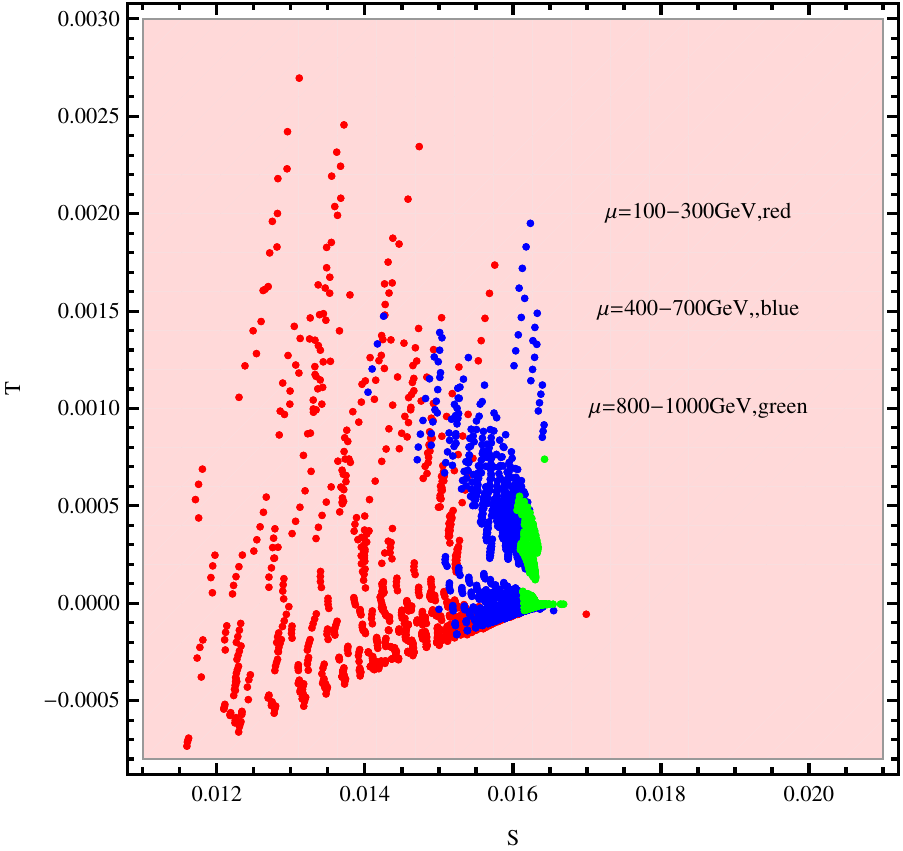}
\end{minipage}%
\caption{Contribution to EWPTs in Higgs sector,
with {\it left} panel for the MSSM with $\tan\beta=10$ and {\it right} panel for the NMSSM with $\tan\beta=\sqrt{7}$.
Constraint $m_{H^{\pm}}\geq 300$ GeV has been imposed.
Red and blue region represents deviation to the SM expectation at 68\% and 95\% CL, respectively.
In the {\it left} panel, red and blue points are covered by the green ones,
and they are inside the 68\% band.}
{\label{higgs}}
\end{figure}

In contrast to the {\it right} panel for the NMSSM,
there is only positive $T$ of order of $10^{-4}\sim 10^{-3}$ for the MSSM with $m_{H^{\pm}}\geq 300$ GeV.
It attributes to the Higgs scalar mass spectrum near the region of decoupling limit. 
The analytic expression of $T$ can be found in \cite{Zheng1}.
Under the decoupling limit one obtains
\footnote{We thank the referee for pointing out this approximation.
The errors in the previous numerical program have been corrected, 
which then reproduces $T$ as desired.},
\begin{eqnarray}
T\simeq \frac{m^{2}_{W}-m^{2}_{Z}\sin^{2}(2\beta)}{48\pi s^{2}_{W}m^{2}_{A}}+\mathcal{O}(m^{4}_{W}/m^{4}_{A}).
\end{eqnarray}
Here $s_{W}$ denotes the weak mixing angle.
Despite the differences between the MSSM and NMSSM mentioned above, 
the contributions to $S$ and $T$ in the Higgs sector are actually small in both cases 
if compared with the neutralino-chargino sector and the stop sector 
that we are going to discuss.

\begin{figure}
\centering
\begin{minipage}[b]{0.4\textwidth}
\centering
\includegraphics[width=2.5in]{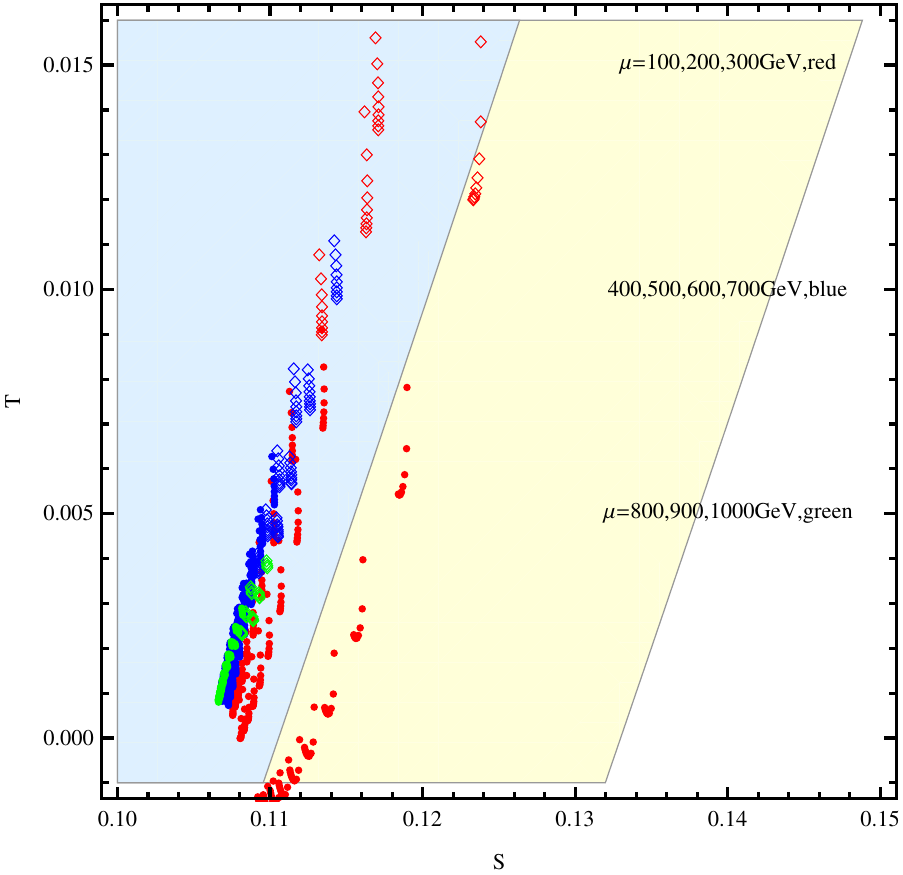}
\end{minipage}%
\centering
\begin{minipage}[b]{0.4\textwidth}
\centering
\includegraphics[width=2.5in]{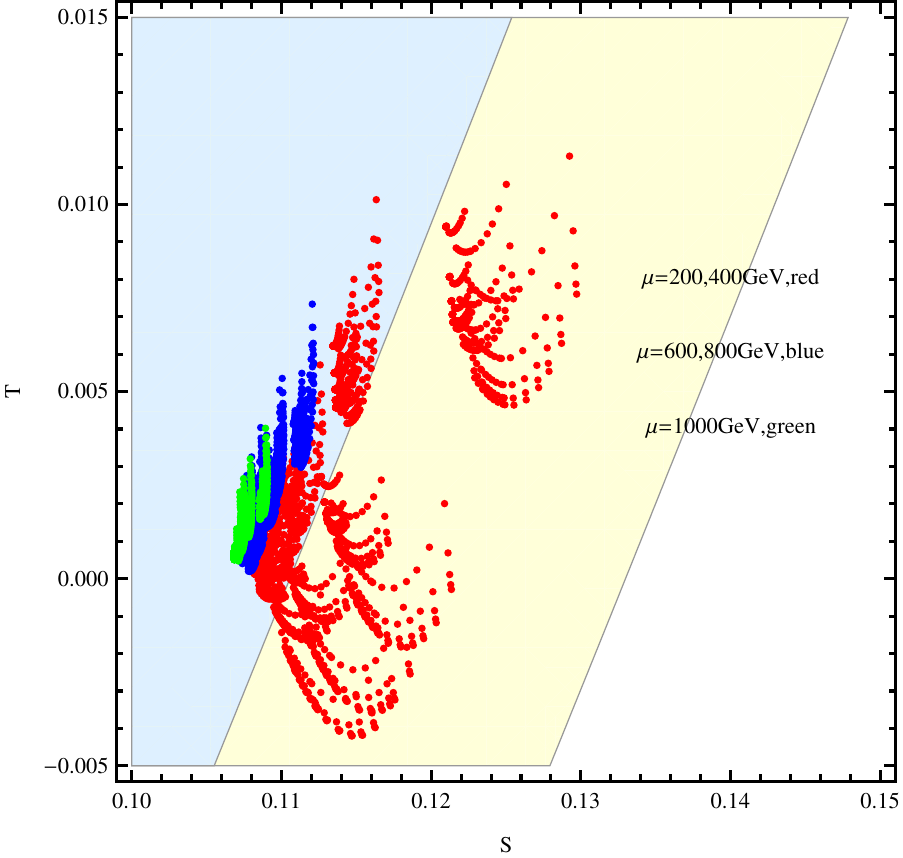}
\end{minipage}%
\caption{Contribution to EWPTs in the neutralino-chargino sector,
with {\it left} panel for the MSSM with $\tan\beta=10$ and {\it right} panel for the NMSSM with $\tan\beta=\sqrt{7}$.
The constraint that lighter chargino mass $\geq 103$ GeV has been imposed.
Blue and yellow background corresponds to deviation to the SM expectation at 95\% and 99\% CL, respectively.
In the {\it left} panel, for different $\mu$ value diamond points refer to light chargino mass beneath 103 GeV, which is excluded by data of LEP II.
In the {\it right} panel, no light chargino below LEP II  mass bound appears.
In either case, $S$ and $T$ approaches to $\sim +0.13$  and $\sim 0.10$ at most, respectively.
}
{\label{neutralinochargino}}
\end{figure}

\subsection{Neutralino-Chargino Sector}
We present contribution to EWPTs in the neutralino-chargino sector in Fig.\ref{neutralinochargino},
with {\it left} and {\it right} panel corresponding to the MSSM and NMSSM, respectively.
The blue and yellow background corresponds to deviation to the SM expectation at 95\% and 99\% CL, respectively.
In the {\it left} panel,  {\it diamond} points refer to light chargino mass beneath 103 GeV, 
which is excluded by data of LEP II.
A light chargino is a result of either small $\mu$ value and /or large mass splitting between the two charginos. 
An increase of $\mu$ corresponds to heavier chargino masses.
In the {\it right} panel, no light chargino below LEP II  mass bound appears.
Each panel of Fig.\ref{neutralinochargino} shows that large deviations are present when $\mu$ saturates its lower bound $\sim 100$ GeV. 
This follows from the fact that a small $\mu$ leads to a relatively larger mass splitting between two charginos.

The analytic expression for $S$ and $T$ in this sector is presented in \cite{Zheng2}.
Numerical calculation shows that $S$ and $T$ approaches to $\sim 0.13$  and $\sim 0.10$ at most, respectively.
Our scan shows that the numerical regions of $S$ and $T$ correspond to,
\begin{eqnarray}{\label{nccontribution}}
\mathbf{MSSM} &:&  0.105 \leq S_{NC} \leq 0.125, ~~~ -0.001 \leq T_{NC}\leq 0.015,\nonumber\\
\mathbf{NMSSM} &:&  0.105 \leq S_{NC} \leq 0.13, ~~-0.005 \leq T_{NC}\leq 0.01.
\end{eqnarray}
Contrary to the MSSM, Fig.\ref{neutralinochargino} clearly shows that, 
for a given value of $\mu$, the NMSSM receives a larger contribution.
This can be ascribed to a relatively larger mass splitting present in the NMSSM.

\begin{figure}
\centering
\begin{minipage}[b]{0.4\textwidth}
\centering
\includegraphics[width=2.5in]{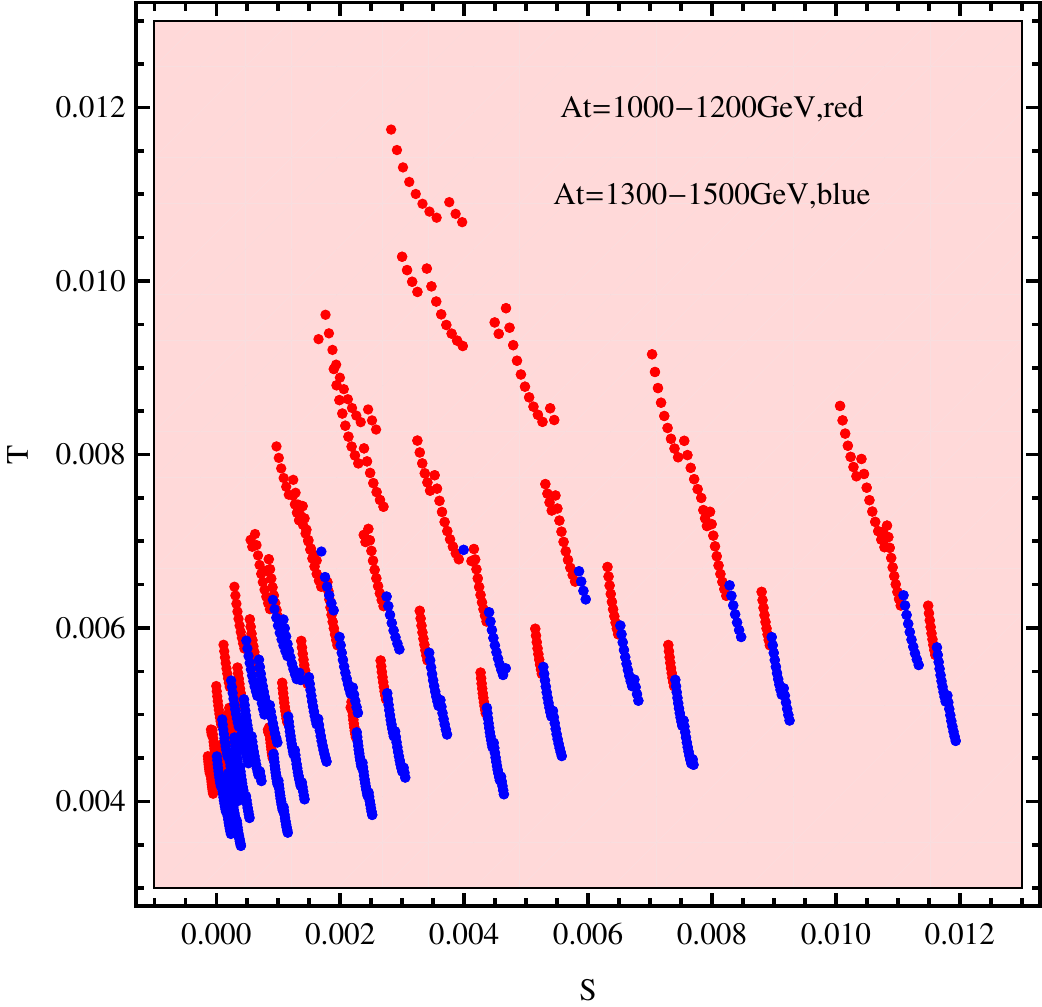}
\end{minipage}%
\centering
\begin{minipage}[b]{0.4\textwidth}
\centering
\includegraphics[width=2.5in]{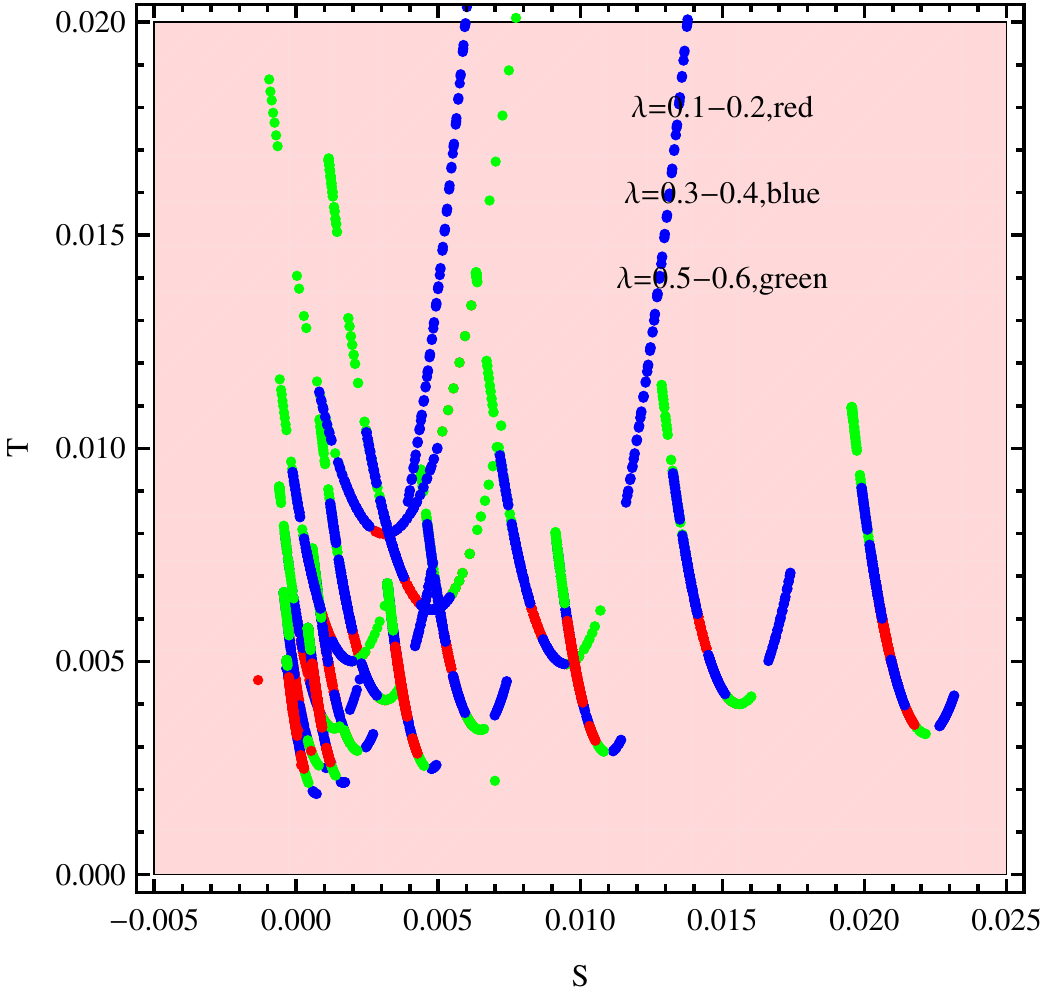}
\end{minipage}%
\caption{Contribution to EWPTs in the third-generation squark sector,
with {\it left} panel for the MSSM with $\tan\beta=10$ and {\it right} panel for the NMSSM with $\tan\beta=\sqrt{7}$.
Here $m_{\tilde{t}_{L}}\simeq m_{\tilde{b}_{L}}$ and $m_{\tilde{t}_{R}}\simeq m_{\tilde{b}_{R}}$ have been adopted,
which implies that contribution to EWPTs is totally induced by $A_t$ term.
The constraint on Higgs mass in the range $125\pm 1$ GeV has been imposed.
Red background corresponds to deviation to the SM expectation at 68\% CL.
$S$ approaches to $0.012$ in the {\it left} panel, 
whereas it is related to the magnitude of $\lambda$ in the {\it right} panel.
}
{\label{squark}}
\end{figure}

\subsection{The third-generation Squark Sector}
Fig.\ref{squark} shows the contribution to EWPTs in the third-generation squark sector,
with {\it left} and {\it right} panel corresponding to the MSSM and NMSSM, respectively.
Red background corresponds to deviation to the SM expectation at 68\% CL	.
In the {\it left} panel, points with different colors refer to different values of $A_t$. 
The figure shows that $S$ goes to 0.01 as $A_t$ increases.
Our scan shows the numerical region of $S$ and $T$,
\begin{eqnarray}{\label{stopcontribution}}
\mathbf{MSSM} &:&  0 \leq S_{q_{3}} \leq 0.012, ~~~ 0 \leq T_{q_{3}}\leq 0.012.
\end{eqnarray}

Differently from the MSSM, 
the $S$ parameter in the NMSSM can be negative, while $T$ remains always positive.
The reason can be partially understood as follows.
As shown in \cite{Zheng1} the analytic expressions for $S$ ad $T$ in the case of significant mixing 
between the left- and righ-handed squarks are rather complicated.
It is not obvious to determine the sign of $S$ and $T$ in this case.
But $S$ and $T$ dramatically reduce to simple form for small mixing \cite{Zheng3},
\begin{eqnarray}{\label{reduction}}
S\simeq -\frac{1}{12\pi} \ln(x), ~~~~~~~
T \simeq \frac{3m^{2}_{\tilde{d}_{3}}}{16\pi s^{2}_{W}m^{2}_{W}} h(x).
\end{eqnarray}
where $x=m^{2}_{\tilde{u}_{3}}/m^{2}_{\tilde{d}_{3}}$ and $h(x)=1+x-\frac{2x}{x-1}\ln(x)$.
Unlike the MSSM, the NMSSM includes the case without mixing effect, 
so Eq.(\ref{reduction}) should explain part of points in the {\it right} panel.
Indeed, $S$ in Eq.(\ref{reduction}) has either sign 
and the positive $h(x)$ explains the positivity of $T$.

In the {\it right} panel points in different colors refer to different regions of $\lambda$.
Naively, the parameter $\lambda$ does not directly contribute to the EWPTs in this sector.
Nevertheless, it controls the magnitude of the loop correction to the Higgs mass, 
which in turn determines the stop masses.
Moreover, it measures how the MSSM deviates from the NMSSM.
Given the same $\tan\beta$, the NMSSM approaches to the MSSM as $\lambda\rightarrow 0$.
This can be verified from the similarity between the patterns of the red points in the {\it right } panel and the points in the {\it left} panel.

\begin{figure}
\centering
\begin{minipage}[b]{0.4\textwidth}
\centering
\includegraphics[width=2.5in]{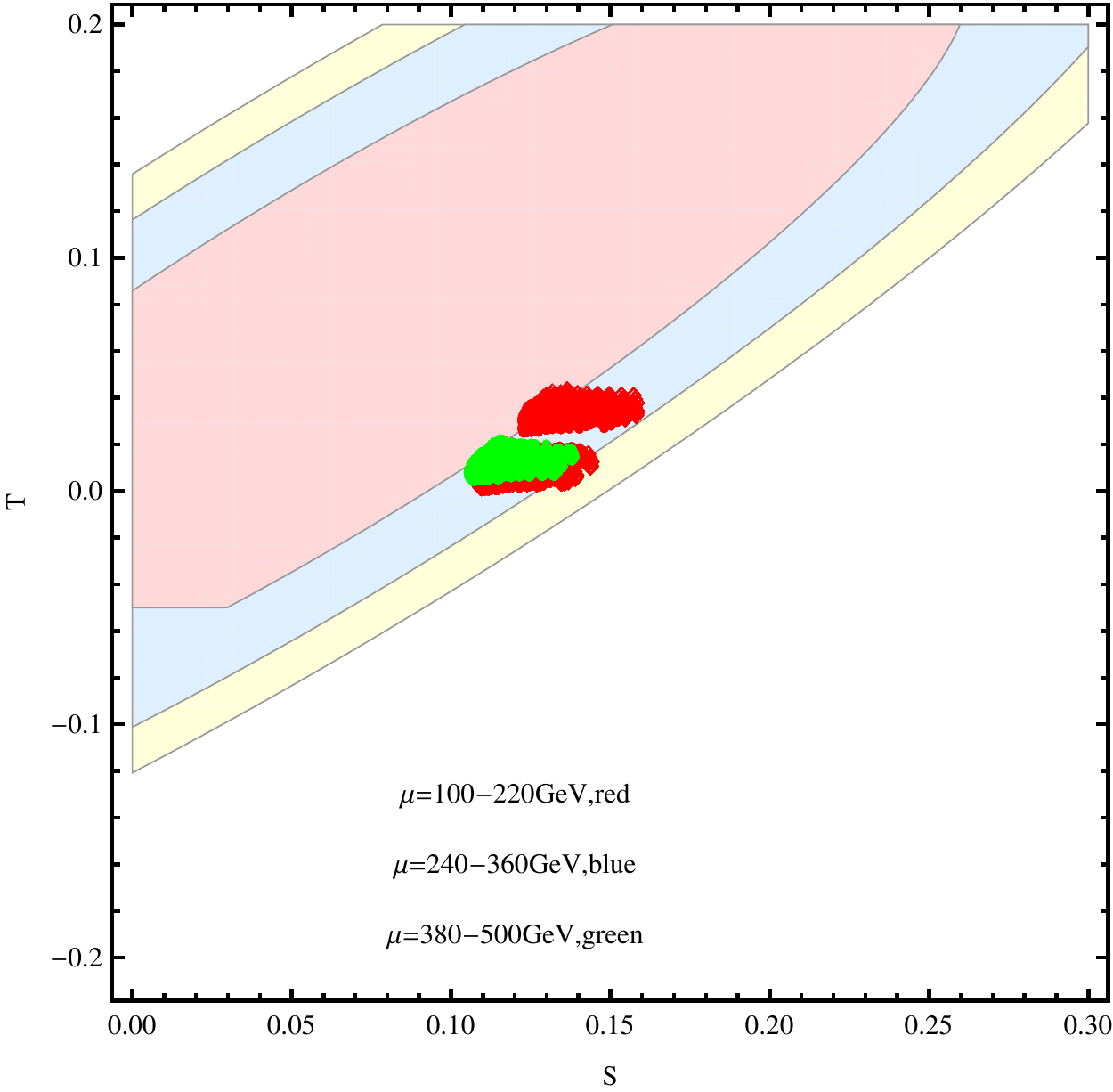}
\end{minipage}%
\centering
\begin{minipage}[b]{0.4\textwidth}
\centering
\includegraphics[width=2.5in]{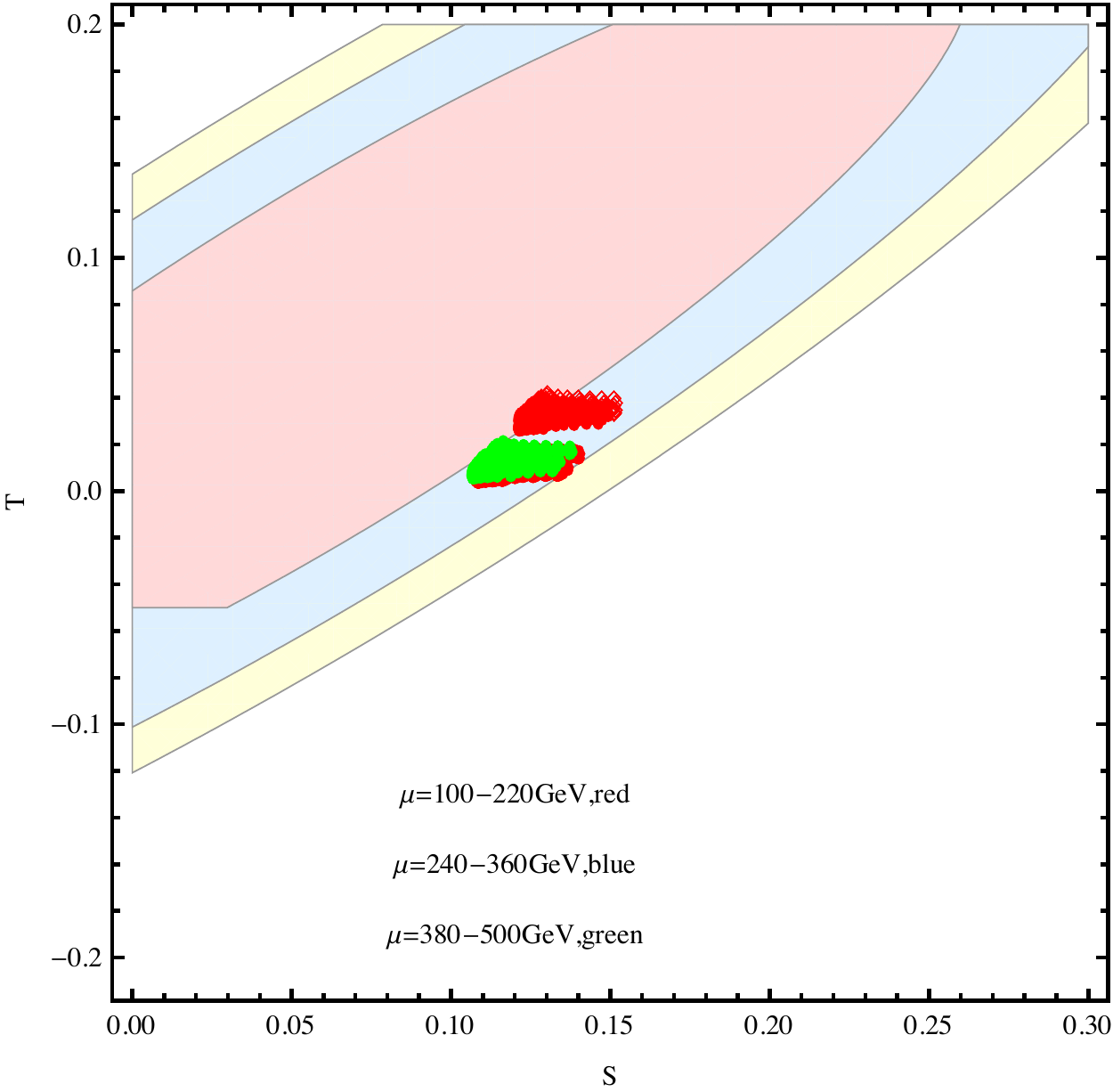}
\end{minipage}%
\caption{Total contribution to EWPTs for the MSSM with $\tan\beta=10$ (left) and  $\tan\beta=20$ (right), respectively.
Natural MSSM which satisfies light chargino mass bound, charged Higgs mass bound
and $126\pm 1$ GeV Higgs mass constraint is nearly excluded at 68\% CL but still consistent at 95\% CL. }
{\label{totalmssm}}
\end{figure}

We would like to mention that in the {\it right} panel points with the same color are widely distributed.
It is a consequence of two facts.
The first one is that the overall magnitude of $S$ ($T$) is actually small.
Second, the fit to Higgs mass is more sensitive to $\lambda$ than the ratio $A_{t}/m_{\tilde{t}}$,
where $\lambda$ and this ratio measures the tree-level and one-loop contribution to the Higgs mass squared, respectively.
Therefore, the ratio $A_{t}/m_{\tilde{t}}$, which measures the mass splitting in stop masses, 
is not necessary to satisfy the maximal mixing condition $X_{t}\simeq \sqrt{6} m_{\tilde{t}}$, 
where $X_{t}=A_{t}-\mu\cot\beta$ and $m_{\tilde{t}}$ the average stop mass.
It can thus be either larger or smaller than that in the MSSM, explaining the wide distribution.

\subsection{Total Contribution}
Fig.\ref{totalmssm} shows the total contributions to EWPTs for the MSSM,
with the {\it left } and {\it right} panel corresponding to $\tan\beta=10$ and $\tan\beta=20$, respectively.
We find that natural MSSM, which is consistent with light chargino mass bound, 
charged Higgs mass bound, and reproduces the observed Higgs mass, is shown to be consistent at 95\% CL.
The differences between the two are very small, 
meaning that the total contribution in the large $\tan\beta$ region is not sensitive to $\tan\beta$.
In Fig.\ref{totalmssm}  $m_{\tilde{t}_{L}}\simeq m_{\tilde{b}_{L}}$ 
and $m_{\tilde{t}_{R}}\simeq m_{\tilde{b}_{R}}$ have been adopted,
which implies that the contribution to the EWPTs is totally controlled by $A_t$ term.
Relaxing these degenerate conditions will lead to larger contributions to EWPTs.

\begin{figure}[h!]
\centering
\begin{minipage}[b]{0.4\textwidth}
\centering
\includegraphics[width=2.5in]{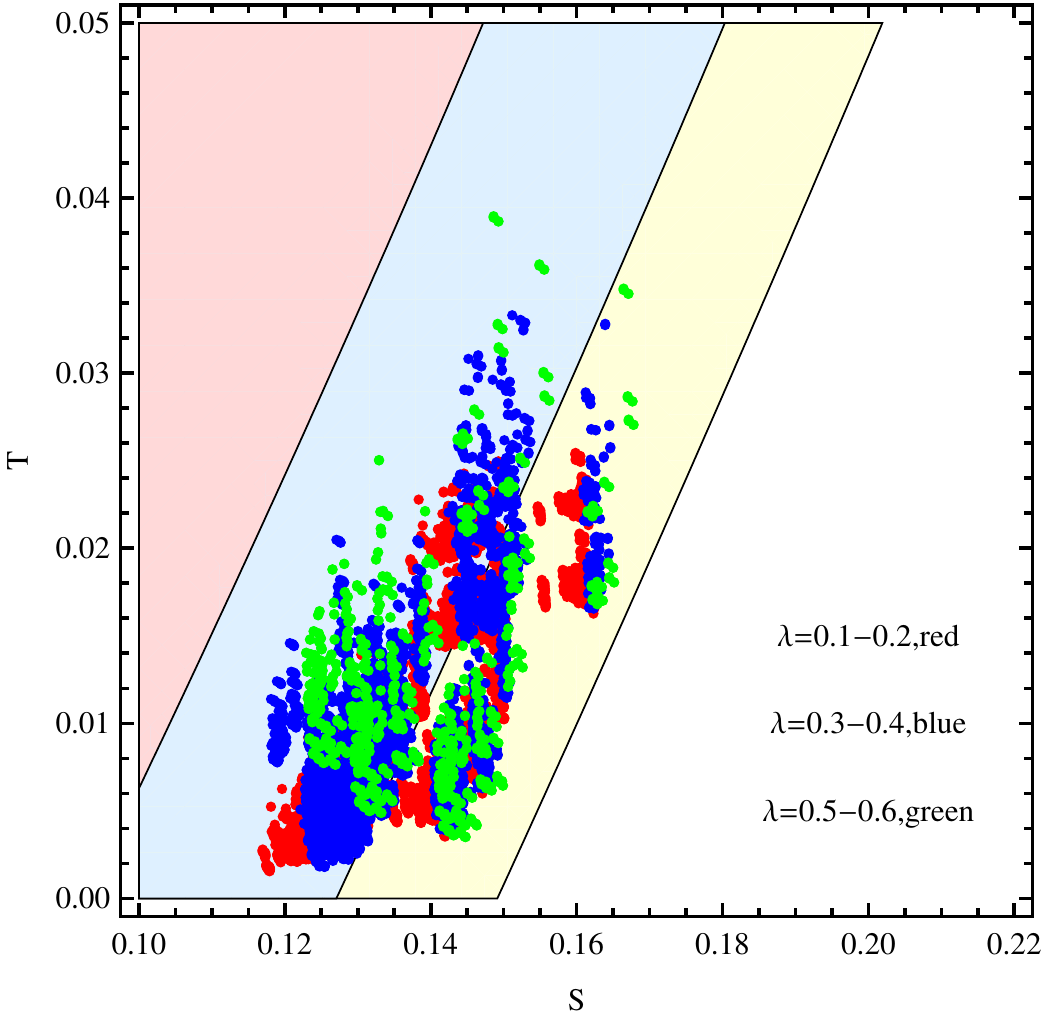}
\end{minipage}%
\centering
\begin{minipage}[b]{0.4\textwidth}
\centering
\includegraphics[width=2.5in]{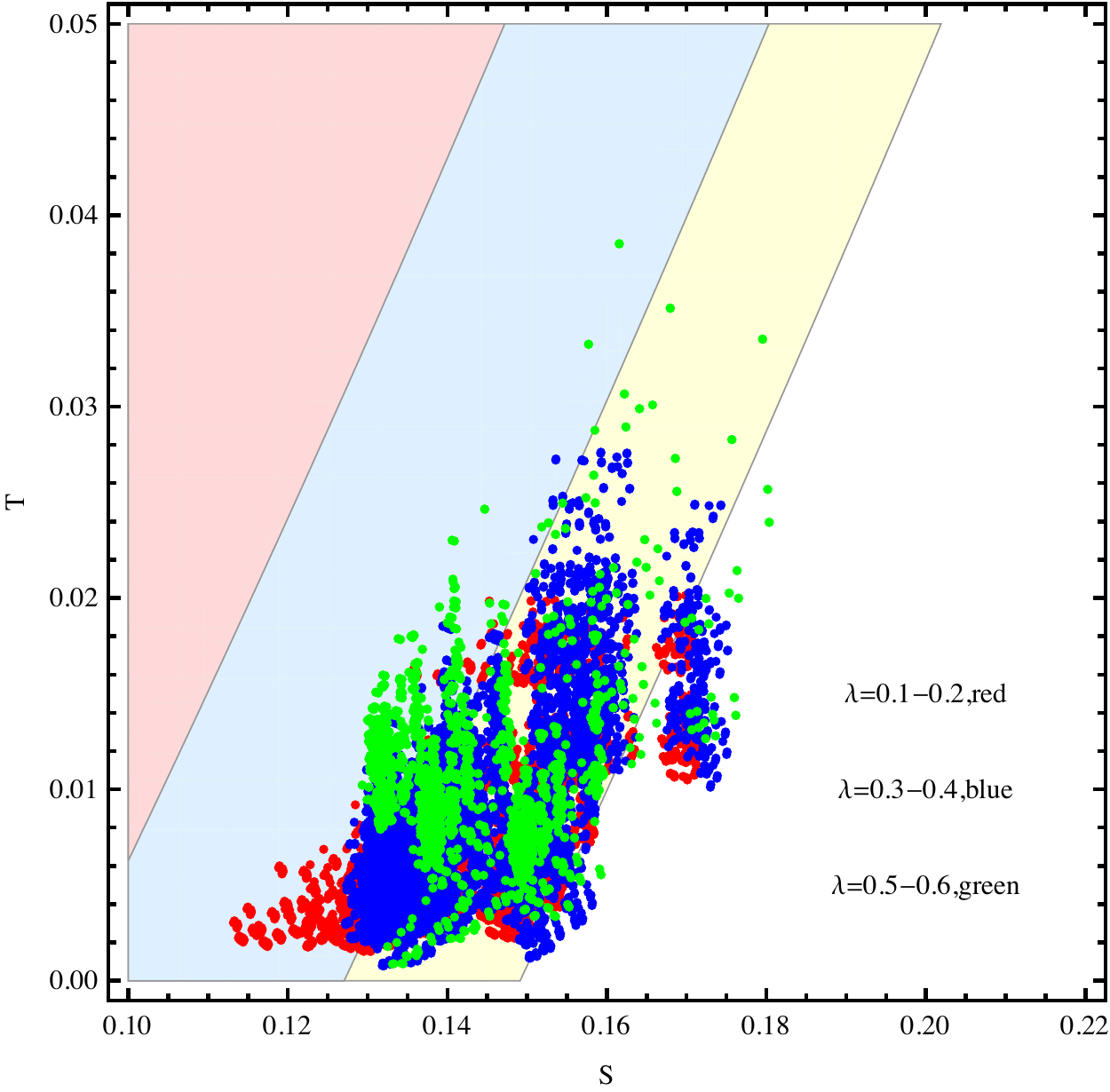}
\end{minipage}%
\caption{Total contribution to EWPTs characterized by $\lambda$ for the NMSSM 
with $\tan\beta=\sqrt{7}$ ({\it left}) and $\tan\beta=2$ ({\it right}), respectively.
Nearly degenerate conditions $m_{\tilde{t}_{L}}\simeq m_{\tilde{b}_{L}}$ 
and $m_{\tilde{t}_{R}}\simeq m_{\tilde{b}_{R}}$ have been adpoted, similar to the MSSM.
Natural NMSSM, which satisfies all the constraints as the same as the MSSM,
is shown to be excluded by EWPTs at 68\%  CL but consistent at 99\% CL for $\lambda\leq 0.6$.
}
{\label{totalnmssm1}}
\end{figure}

We show the total contributions to EWPTs for the NMSSM in Fig.\ref{totalnmssm1},
with the {\it left } and {\it right} panel corresponding to $\tan\beta=2$ and $\tan\beta=\sqrt{7}$, respectively.
We find that natural NMSSM, which is consistent with the light chargino mass bound, 
charged Higgs mass bound, and and reproduces the observed Higgs mass, 
is shown to be excluded at 68\% CL but still consistent at 99\% CL for $\lambda\leq 0.6$.
As $\lambda$ increases, the differences between the two panels are more obvious.
This figure also shows that all the points tend to move outside the contour when $\tan\beta$ is far way from its central value.
So, we expect that in cases corresponding to either $\tan\beta\simeq 1$ or $\tan\beta >>\mathcal{O}(1)$ natural NMSSM is excluded at 99\% CL.
Note that in Fig.\ref{totalnmssm1} we have adopted nearly degenerate conditions 
$m_{\tilde{t}_{L}}\simeq m_{\tilde{b}_{L}}$ 
and $m_{\tilde{t}_{R}}\simeq m_{\tilde{b}_{R}}$.
Relaxing these conditions leads to larger contribution to EWPTs,
which will strengthen our conclusion.

\begin{figure}[h!]
\centering
\begin{minipage}[b]{0.7\textwidth}
\centering
\includegraphics[width=4in]{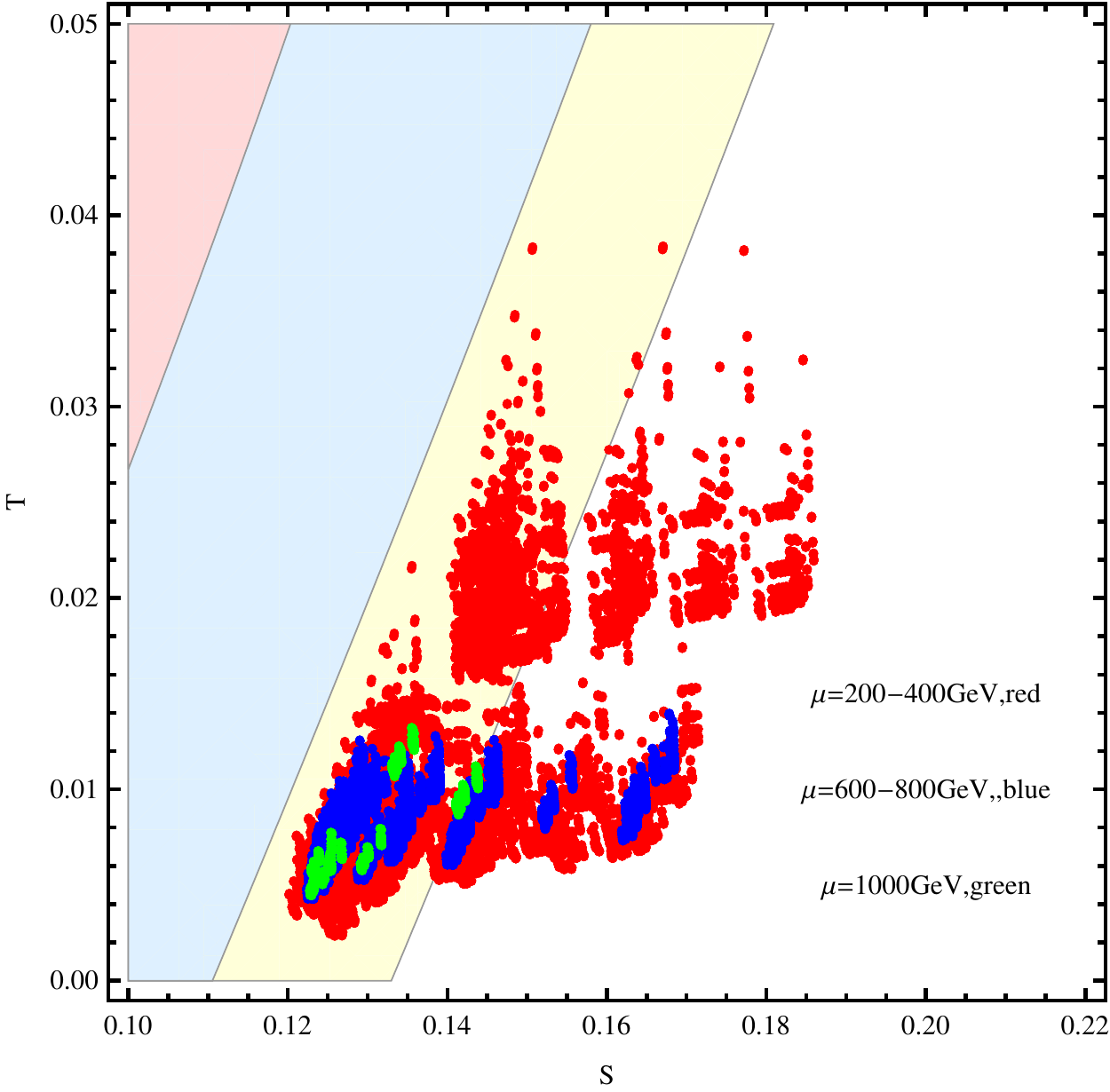}
\end{minipage}%
\caption{Total contribution to EWPTs characterized by $\mu$ for the NMSSM 
with $\tan\beta=\sqrt{7}$. }
{\label{totalnmssm2}}
\end{figure}

In Fig.\ref{totalnmssm2} we show the total contributions to EWPTs for the NMSSM characterized by $\mu$ instead of $\lambda$. It indicates that natural NMSSM is excluded by EWPTs at 68\% CL but still consistent at 99\% CL for $\mu\leq 1$ TeV.

Our results for either the MSSM or the NMSSM are quite insensitive to the soft mass spectrum, 
provided it is of a ``natural'' type.
It is also independent on mechanisms of SUSY breaking.

\section{Conclusions}
In this paper we have explored the natural SUSY spectrum in the light of EWPTs. 
The narrow window for the input parameters allowed both by naturalness and LHC data 
offers a well defined region that can be studied through the EWPTs.
The main results of this study are the following ones.
\begin{itemize}
\item Natural MSSM, 
which satisfies the light chargino mass bound, charged Higgs mass bound and reproduces the observed Higgs mass, 
is excluded at 68\% CL but still consistent at 95\% CL.
\item Natural NMSSM with degenerate conditions, 
which satisfies the same constraints as the MSSM, 
is excluded at 68\%  CL but still consistent at 99\% CL for either $\mu\leq 1$ TeV or $\lambda\leq 0.6$.
\end{itemize}
These observations hold in any SUSY model whose mass spectrum is similar to that of natural SUSY at the electroweak scale.

There are some interesting directions along this line.
First, our analysis can be generalized to SUSY model with  $\lambda$ larger than what is considered here, 
which is called $\lambda$-SUSY.
Second, it is also of interest to relax the degenerate conditions,
and see how tightly the conclusion will be strengthened.
Finally, the combination of EWPTs and precise measurement on Higgs couplings 
may lead to more solid claims for natural SUSY.
\\

~~~~~~~~~~~~~~~~~~~~~~~~~~~~~~~~~~~~~~
${\it\mathbf{ Acknowledgement}}$\\
The authors thank the referee for valuable suggestions.
The work is supported in part by Natural Science Foundation of China 
under grant No.11247031 and 11405015.

\end{document}